\documentclass[aps, pra, preprint]{revtex4-1}
\usepackage{graphicx}
\usepackage{amsmath}
\usepackage{amssymb}
\usepackage{dsfont}
\usepackage[utf8]{inputenc}
\usepackage{geometry}
\usepackage{placeins}
\def\nn{\nonumber}
\def\beq{\begin{equation}}
\def\eeq{\end{equation}}
\def\beqna{\begin{eqnarray}}
\def\eeqna{\end{eqnarray}}
\def\bea{\begin{array}}
\def\ea{\end{array}}

\begin{document}
\title{Rabi oscillations in two-level systems beyond the rotating-wave 
approximation}
\author{Adriano A. Batista}
\email{adriano@df.ufcg.edu.br}
\affiliation{
Departamento de F\'{\i}sica\\
Universidade Federal de Campina Grande\\
Campina Grande-PB\\
CEP: 58109-970\\
Brazil}
\date{\today}
\begin{abstract}
Here we use perturbation techniques based on the averaging method
to investigate Rabi oscillations in cw and pulse-driven two-level systems 
(TLS's).
By going beyond the rotating-wave approximation, especifically
to second-order in perturbation, we obtain  the Bloch-Siegert shift
of the TLS resonant frequency, in which the resonant frequency
increases with the driving field amplitude.
This frequency shift implies that short resonant $\pi$-pulses 
in which the Rabi frequency is  approximately 40\% or higher of the 
transition frequency do not achieve complete inversion in TLS's.
Hence, guided by analytical results based on the averaging technique, we
propose two methods for obtaining
population inversions in the TLS driven by short $\pi$-pulses:
one with chirping and the other with pulse shaping and
near resonance blue-shifted detuning.
Both methods minimize dephasing due to the Bloch-Siegert shift,
reduce the dependance of the excitation of the TLS on the pulse phase,
and are very effective in achieving complete population inversions.

\end{abstract}
\maketitle
\section{Introduction}
% Background of the present work
Here we investigate the coherent, cw or pulsed, excitation of two-level
systems (TLS's) in quantum mechanics, under the semiclassical approximation.
Our main purpose is to develop approximate analytical methods with better
precision than the rotating-wave approximation (RWA) that will help us in
designing more accurate and faster coherent control schemes than what is
currently available in the literature. 

%Present results
We use the perturbative method known as the averaging method \cite{Guck83,
verh96} to go one step beyond RWA.
Usually this method is applied to periodically-driven dynamical systems with 
two very different time scales (one fast and the other slow), although, as we
will see, even when the time scales are not very far apart the method is still
very useful.
Furthermore, this method has the added advantage that it can provide
efficiently both transient and stationary approximate solutions, unlike the
usual time-dependent perturbative methods of quantum mechanics.
The proposed approach is considerably simpler than to implement than
Floquet theory \cite{shirley1965solution, sambe1973steady}, which technically
is not applicable to TLS's excited by pulses.
By implementing our method, we obtained  the first correction to the resonant
frequency of the TLS as the pump amplitude is increased, an effect known as
the Bloch-Siegert shift \cite{bloch1940magnetic}.

Quantum control of population transfer in TLS's is a topic of intense current
theoretical and experimental investigations \cite{bergmann1998coherent,
wollenhaupt2006quantum, trallero2007transition, conover2011effects}, 
mostly with the purpose of implementing quantum gates in various physical
systems for quantum computing. 
Hence, it is desirable that laser pulses used in the realization
of quantum gates be as short in time as possible so that they conform with
basic theoretical models of quantum computing \cite{nielsen2010quantum}.
Furthermore, such short pulses reduce the adverse effect of decoherence, 
which then reduce the amount of errors involved in computation with qubits.
On the other hand, the fields become stronger, the Fourier spectrum widens, and
the control becomes more difficult, with the breaking up of the RWA and,
consequently, of the area theorem \cite{mccall67, mccall67b, allen87}, which
depends on the RWA.
In such situations, the effectiveness of short and high-amplitude
$\pi$-pulses depends on the phase of the pulse as well.
Hence, in order to diminish such difficulties, we designed $\pi$-pulses, either with
chirp in the carrier frequency or with pulse shaping with near resonance
blue-shifted carrier frequency that minimizes the dephasing effects due
to Bloch-Siegert shift. 
Analytical and numerical results of this approach of pulse chirping or shaping
are presented with $\pi$-pulse-driven TLS's that yield considerably deeper
interlevel population transfers and have almost no dependance on the phase of
the pulses.
This adds to recent investigations on quantum control of TLS's that try
to achieve maximum population transfer either with chirping, which mostly use
linearly-chirped pulses
\cite{cao1998molecular,zamith2001observation,jha2010analytical} 
or with pulse shaping \cite{boradjiev2013control}.

% Brief description of the following sections
This paper is organized as follows.
In Sec.~\ref{sec:RWA}, we review the results of RWA for the Schrödinger equation
for the ac-driven TLS and present a naive perturbation method for
tentatively going beyond  RWA.
In Sec.~\ref{sec:avg2}, we obtain the 2nd-order averaging method correction to 
the RWA time-evolution.
In Sec.~\ref{sec:dm}, we obtain the density-matrix equations of motion in the
2nd-order averaging approximation.
In Sec.~\ref{sec:qcontrol}, we apply our theoretical results in the design of 
effective $\pi$-pulses.
In Sec.~\ref{sec:conclusion}, we  discuss possible
applications of our main results and draw our conclusions.
\section{Rabi oscillations in a two-level system}
\label{sec:RWA}
The problem of coherent control of TLS's has a long and intense history, dating back to the 30's.
It started with the study of the dynamics of magnetization of nuclei with spin
1/2 interacting with a magnetic field $\vec B(t)=\vec B_{dc}+\vec
B_{ac}\cos(\omega t)$, in which there is an angle between the dc and
ac components of the magnetic field. 
This was originally investigated and solved in the RWA, in a semiclassical
approach, by I.  Rabi in Ref. \cite{rabi1937space}.
He found that, in resonance, there is a precession of
the magnetization around the ac field, which became known as Rabi cycle, at a
frequency $g\mu_B B_{ac}/\hbar$, known as Rabi frequency, in which $g$
is the gyromagnetic constant and $\mu_B$ is the Bohr magneton.
This system has been found to have a time evolution equivalent to 
that of a TLS (atom, molecule, quantum dot, exciton,
etc) interacting with a single mode of the electromagnetic field inside an
optical cavity.

This generic Hamiltonian for this dynamical system can be written as
\beq
\mathcal{H}(t)=E_1|1\rangle\langle 1|+E_2|2\rangle\langle2|+2\hbar \Omega_0\cos(\omega t)\left[|1\rangle\langle2|+|2\rangle\langle1|\right].
\eeq
The wave function in the two-level approximation is given by 
\[
|\psi(t)\rangle=C_1(t)|1\rangle+C_2(t)|2\rangle.
\]
Hence, the corresponding Schr\"odinger equation is
\begin{subequations}
\begin{align}
\imath\hbar \dot C_1 &= E_1C_1+2\hbar \Omega_0\cos(\omega t)C_2,\\
\imath\hbar \dot C_2 &= E_2 C_2+2\hbar \Omega_0\cos(\omega t)C_1.
\end{align}
\label{eq:schrodinger}
\end{subequations}

With the transformations $C_1(t)=e^{-\frac{\imath E_1}{\hbar} t} c_1(t)$ and 
$C_2(t)=e^{-\frac{\imath E_2}{\hbar} t}c_2(t)$, we obtain the interaction
picture dynamics, which is given by
\begin{subequations}
\begin{align}
    \imath\dot c_1 &= 2\Omega_0e^{-i\omega_0 t}\cos(\omega t)c_2
    =\Omega_0e^{\imath\delta t}\left(1+e^{-2\imath\omega t}\right)c_2,\\
\imath \dot c_2 &= 2\Omega_0e^{i\omega_0 t}\cos(\omega t)c_1=\Omega_0e^{-\imath\delta t}\left(1+e^{2\imath\omega t}\right)c_1,
\end{align}
\label{eq:2level}
\end{subequations}
where $\omega_0=\frac{E_2-E_1}{\hbar}$ is the transition frequency
and $\delta=\omega-\omega_0$ is the detuning.
In general this system would be quasiperiodically driven, 
and thus one could not readily apply the averaging method or Floquet theory.
One fixes this problem with the following transformation 
\begin{subequations}
\begin{align}
    c_1(t) &=e^{\dfrac{\imath\delta t}{2}}b_1(t),\\
    c_2(t) &=e^{-\dfrac{\imath\delta t}{2}}b_2(t), 
\end{align}
\label{c_to_b}
\end{subequations}
we obtain the equations of motion
\begin{subequations}
\begin{align}
    \imath\dot b_1 &= \dfrac{\delta}{2}b_1+\Omega_0\left(1+e^{-2\imath \omega t}\right)b_2,\\
    \imath \dot b_2 &= -\dfrac{\delta}{2} b_2+\Omega_0\left(1+e^{2\imath\omega t}\right)b_1,
\end{align}
\label{eq:twolevel}
\end{subequations}
which are a parametrically-driven dynamical system with periodic coefficients.
Hence, we can now perform the usual RWA and obtain
\begin{align}
    \imath\dot b_1 &= \frac{\delta}{2} b_1+\Omega_0b_2,\\
    \imath \dot b_2 &= -\frac{\delta}{2} b_2+\Omega_0b_1.
\label{eq:RWA}
\end{align}
For this approximation to be valid, we assume $\delta$ and $\Omega_0=\mathcal{O}(\epsilon)$, with $0<\epsilon<<1$. 
With a few algebraic steps one finds
\[
\ddot b_i=-\left(\Omega_0^2+\dfrac{\delta^2}{4}\right)b_i,
\]
where $i=1,2$.
For the initial values $C_1(0)=b_1(0)=1$ and $C_2(0)=b_2(0)=0$, 
which are equivalent to $b_1(0)=1$ and $\dot b_1(0)=-\imath\delta/2$
or to $b_2(0)=0$ and $\dot b_2(0)=-\imath\Omega_0$
we have
\begin{subequations}
\begin{align}
    b_1(t) &= \cos(\Omega t)-\frac{\imath\delta}{2\Omega}\sin(\Omega t),\\
    b_2(t) &= -\frac{\imath\Omega_0}{\Omega}\sin(\Omega t),
\end{align}
\label{eq:rabi}
\end{subequations}
where $\Omega=\sqrt{\Omega_0^2+\frac{\delta^2}{4}}$.
Hence the occupation probabilities in each level are
\begin{subequations}
\begin{align}
\left|C_1(t)\right|^2 &=\cos^2(\Omega t)+\frac{\delta^2}{4\Omega^2}\sin^2(\Omega t),\\
\left|C_2(t)\right|^2 &=\frac{\Omega_0^2}{\Omega^2}\sin^2(\Omega t).
\end{align}
\label{eq:rabiPop}
\end{subequations}

\begin{figure}[h]
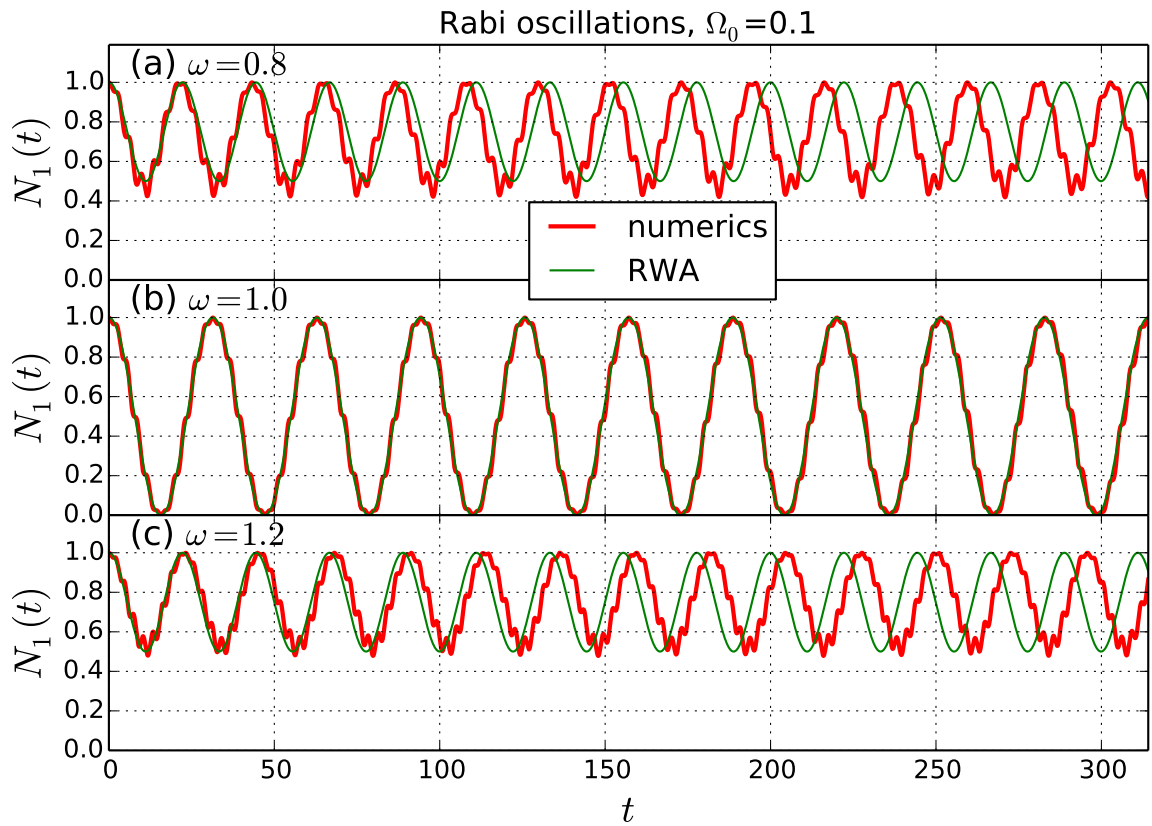

    \centerline{\includegraphics[scale=0.8]{{{rabiOscRWA_Omega0.1}.pdf}}}
        \caption{Comparison of the ground state population time evolution as
        predicted by the RWA in Eq.~\eqref{eq:rabiPop} and the result given by
        numerical integration of Eqs.~\eqref{eq:2level}. 
        In our units $\omega_0=1$.}
        \label{fig:Rabi_flopRWAa}
\end{figure}

\begin{figure}[h]
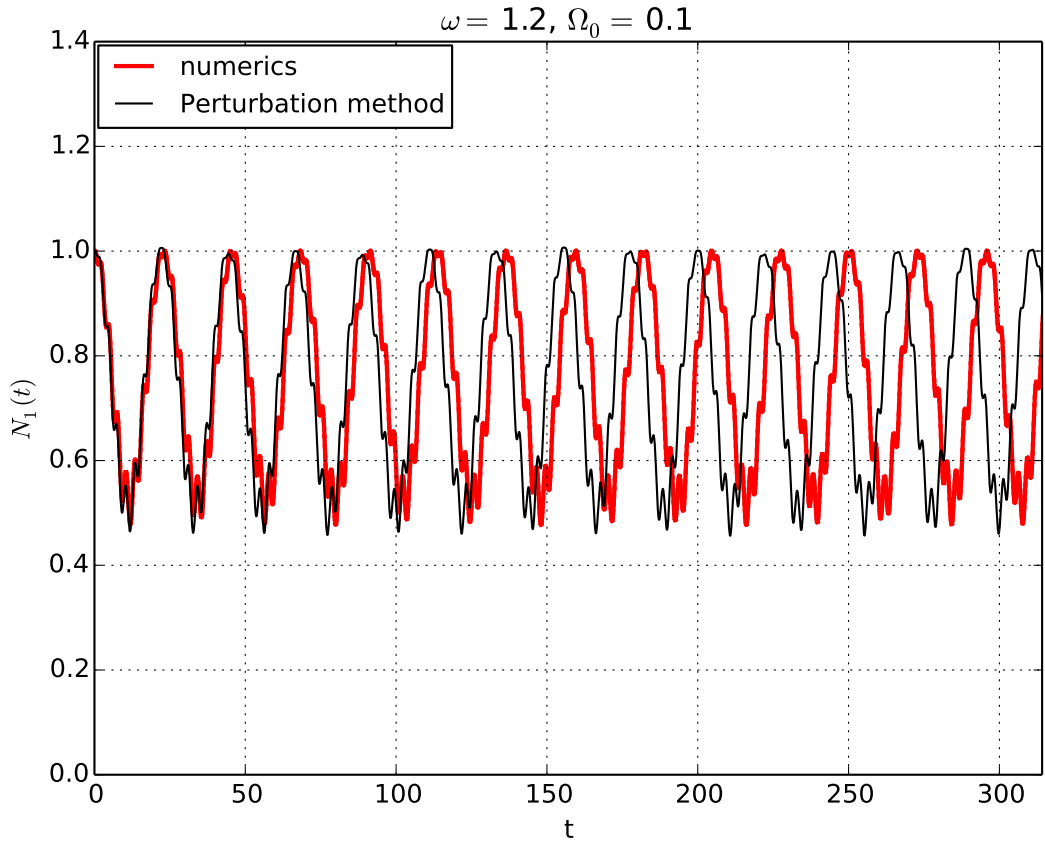

    \centerline{\includegraphics[scale=0.8]{{{rabiOscOmega00.1omega1.2}.pdf}}}
        \caption{Comparison of the ground state population time evolution as
        predicted by the naive perturbative method of Eq.~\eqref{eq:naivePerturb}
        and the result given by numerical integration of
        Eqs.~\eqref{eq:2level}. 
        Note that there is still the same problem of dephasing seen in
        Fig.~\eqref{fig:Rabi_flopRWAa}, showing that this perturbative
        approach does not improve on the RWA.}
        \label{fig:naivePerturb}
\end{figure}

In Fig.~\ref{fig:Rabi_flopRWAa}, we compare the ground state occupation
probability (population) evolution from the numerical integration of
Eq.\eqref{eq:schrodinger} with the corresponding population evolution given by
RWA in Eqs.~\eqref{eq:rabiPop}.
The numerical integrator used was Python's Scipy odeint with a time step of $2^{-10}$.
One sees that there is considerable dephasing between the RWA and numerical
integration results.
In order to reduce this type of problem, one has to go beyond the RWA. 
One could attempt to achieve that by simply replacing the RWA solution,
given in Eqs.\eqref{eq:rabi}, and using the transformation \eqref{c_to_b}
backwards, in the right hand of Eq.\eqref{eq:2level}.
We obtain
\begin{subequations}
\begin{align}
    c_1(t)&=1-2\imath\Omega_0\int_0^te^{-i\omega_0 s}\cos(\omega s)e^{-i\delta s/2}b_2(s)ds\\
    &=1-\frac{2\Omega_0^2}{\Omega}\int_0^t e^{-i(\omega_0 +\delta/2)s}\cos(\omega s)\sin(\Omega s) ds\\
&=1-\frac{\Omega_0^2}{\Omega}\int_0^t e^{-i(\omega_0 +\delta/2)s}
\left\{\sin[(\omega+\Omega)s]+\sin[(\Omega-\omega)s]\right\}ds.
\end{align}
\label{eq:naivePerturb}
\end{subequations}
We need to find the integrals
\begin{align}
 &   \int_0^t e^{-i(\omega_0+\delta/2)s} \sin[(\Omega+\omega)s]ds=
-\frac{e^{i(\Omega +\delta/2)t}-1}{2\Omega+\delta}\nn\\
&-\frac{e^{-i(\Omega+2\omega-\delta/2)t}-1}{2\Omega+4\omega-\delta},\\
&\int_0^t e^{-i(\omega_0 +\delta/2)s}\sin[(\Omega-\omega)s]ds=
\frac{-e^{i(\Omega-2\omega+\delta/2)t}+1}{2\Omega -4\omega+\delta}\nn\\
&-\frac{e^{-i(\Omega -\delta/2)t}-1}{2\Omega-\delta}.
\end{align}
This simple procedure fails to achieve more accurate results than the RWA,
though, as one can see in Fig.~\ref{fig:naivePerturb}, since it does not
decrease the dephasing problem of the RWA.
Another approach has to be devised in order to give a better approximation than
the RWA to the Rabi oscillations without too much burden.
Since the RWA is equivalent to the averaging method in the 1st-order
approximation, a natural candidate to achieve better results than the RWA is
the averaging method to second order.
That is what is accomplished in the next section.
\section{Rabi oscillations using the averaging method to second order}
\label{sec:avg2} 
Although one could apply Floquet theory to solve the system of 
differential equations \eqref{eq:twolevel} exactly, such as was performed in 
Refs. \cite{shirley1965solution, sambe1973steady}, here, instead, we will 
use the simpler  approach of going just one step further than the RWA in
solving this system approximately using the averaging method to second order
\cite{Guck83, bat00}.

Using the notation of Ref. \cite{bat00}, we have
\[
\tilde f(y, t)=-\imath\Omega_0
\left[
\bea{cc}
0                   & e^{-2\imath\omega t}\\
e^{2\imath\omega t} & 0\\
\ea
\right]
\left[
\bea{c}
y_1\\
y_2
\ea
\right],
\]
with the weakly nonlinear transformation $b(t)=y(t)+\epsilon\mathcal{W}(y, t)$,
with $b(t), y(t)\in\mathds{C}^2$,
where  $\mathcal{W}(y, t)$ obeys the differential equation
\beq
\mathcal{W}_t=\tilde f.
\label{eq:Wt}
\eeq
Hence, upon integration of Eq.~\eqref{eq:Wt}, we obtain
\begin{align}
\epsilon\mathcal{W}(y, t)&=\frac{\Omega_0}{2\omega}
\left[
\bea{cc}
0                   & e^{-2\imath\omega t}\\
-e^{2\imath\omega t} & 0\\
\ea
\right]
\left[
\bea{c}
y_1\\
y_2
\ea
\right]\nn\\
&=\frac{\Omega_0}{2\omega}\left[\bea{c} e^{-2\imath\omega t}y_2\\ -e^{2\imath\omega t}y_1\ea\right],
\label{eq:W}
\end{align}
in which the constants of integration were set to zero.
The Jacobian matrix of $\tilde f$ is
\[
\epsilon D\tilde f=-\imath\Omega_0
\left[
\bea{cc}
0                   & e^{-2\imath\omega t}\\
e^{2\imath\omega t} & 0\\
\ea
\right].
\]
With another weakly nonlinear transformation $y(t)=z(t)+\epsilon^2v(z, t)$,
we obtain the second-order averaged system, corresponding to the original
nonautonomous ordinary differential equation (ODE) system of
Eq.~\eqref{eq:twolevel},
\[
\dot z=\epsilon f_0(z)+\epsilon^2\overline{D\tilde f\mathcal{W}},
\]
where the overline denotes the time averaging.
We then find
\[
\epsilon^2\overline{D\tilde f\mathcal{W}}
=-\frac{\imath\Omega_0^2}{2\omega}
\left[
\bea{cc}
-1 & 0\\
0                       & 1
\ea
\right]
\left[
\bea{c}
z_1\\
z_2
\ea
\right].
\]
Consequently, we find
\begin{align}
\left[
\bea{c}
\dot z_1 \\ \dot z_2
\ea
\right]
&=   
-\imath\left[
\bea{cc}
\frac{\delta}{2} &\Omega_0\\
\Omega_0&-\frac{\delta}{2}
\ea\right]
\left[
\bea{c}
z_1 \\ z_2
\ea
\right]
-\frac{\imath\Omega_0^2}{2\omega}
\left[
\bea{cc}
-1 & 0\\
0                       & 1
\ea
\right]
\left[
\bea{c}
z_1 \\ z_2
\ea
\right]\nn\\
&=
-\imath\left[
\bea{cc}
\frac{\tilde\delta}{2} &\Omega_0\\
\Omega_0&-\frac{\tilde\delta}{2}
\ea\right]
\left[
\bea{c}
z_1 \\ z_2
\ea
\right],
\label{eq:rwa2}
\end{align}
where $\tilde\delta=\delta-\frac{\Omega_0^2}{\omega}$.
One sees that the Eqs.\eqref{eq:rwa2} have the same form
as the Eqs.~\eqref{eq:RWA}, with the correction for the detuning.
This correction reduces the dephasing of the usual RWA
for the Rabi oscillations. 
This shift in resonant frequency relative to the RWA resonant
frequency is known as the Bloch-Siegert shift \cite{bloch1940magnetic, shirley1965solution}.
That is not all though, the initial values of the ODE system \eqref{eq:rwa2}
are different from the initial values of Eqs.~\eqref{eq:RWA}.
Hence, that implies that the solution of Eq.~\eqref{eq:rwa2} is given by
\begin{subequations}
\begin{align}
    z_1(t) &= A_1\cos(\tilde\Omega t)+B_1\sin(\tilde\Omega t),\\
    z_2(t) &= A_2\cos(\tilde\Omega t)+B_2\sin(\tilde\Omega t),
\end{align}
\label{eq:rabi_avg2}
\end{subequations}
where the coefficients can be found from the initial values
\begin{align}
    z_1(0) &= A_1=\dfrac{1}{1+\left(\dfrac{\Omega_0}{2\omega}\right)^2}, \\
    z_2(0) &= A_2=\dfrac{\Omega_0/(2\omega)}{1+\left(\dfrac{\Omega_0}{2\omega}\right)^2},\\
    \dot z_1(0)&= \tilde \Omega B_1=-\frac{\imath}{2}\left[\tilde\delta z_1(0)+2\Omega_0z_2(0)\right],\\
    \dot z_2(0)&= \tilde \Omega B_2=-\frac{\imath}{2}\left[ 2\Omega_0z_1(0)-\tilde\delta z_2(0)\right]. 
\end{align}
\section{The density matrix equations of motion}
\label{sec:dm}
In order to investigate one qubit dynamics with depopulation and decoherence
effects, one has to use the density-matrix equations of motion, instead of the
Schrödinger equation.
We start with the following equations
\begin{subequations}
\begin{align}
    \dot \Delta &= -\gamma_1(\Delta -\Delta_0)+4\mbox{Im\,}\sigma F(t),\\
    \dot \sigma &=-\gamma_2\sigma+\imath\omega_{0}\sigma-\imath\Delta F(t),
\end{align}
\label{eq:dm}
\end{subequations}
where $F(t)=2\Omega_0\cos(\omega t+\varphi_0)$, $\Delta$ is the population difference $N_1(t)-N_2(t)$, $\Delta_0$ is the
equilibrium population difference, and $\sigma(t)$ is
the off-diagonal element $\rho_{21}(t)$ of the density matrix in the basis of
the two states $|1\rangle$ and $|2\rangle$.
The parameters $\gamma_1=1/T_1$, the depopulation rate, and $\gamma_2=1/T_2$, the decoherence rate, are dissipation rates obtained from the relaxation-time
approximation.
We now proceed to implement the averaging method with the transformation
$\sigma(t)=e^{\imath\omega t}\tilde\sigma(t)$ to set the Eqs.~\eqref{eq:dm} in
the slowly-varying frame (also known as the interaction representation).
Hence, we obtain
\begin{subequations}
\begin{align}
    \dot \Delta &= -\gamma_1(\Delta -\Delta_0)-2\imath\Omega_0 \left(\tilde\sigma e^{\imath\omega t} -\tilde\sigma^*e^{-\imath\omega t}\right)\left(e^{\imath(\omega t+\varphi_0)}+e^{-\imath(\omega t+\varphi_0)}\right),\\
    \dot{\tilde\sigma} &=-\gamma_2\tilde \sigma-\imath\delta\tilde\sigma
    -\imath\Omega_0\Delta\left[e^{\imath\varphi_0}+e^{-\imath(2\omega t+\varphi_0)}\right],\\
    \dot{\tilde\sigma}^* &=-\gamma_2\tilde \sigma^*+\imath\delta\tilde\sigma^*
    +\imath\Omega_0\Delta\left[e^{-\imath\varphi_0}+e^{\imath(2\omega t+\varphi_0)}\right],
\end{align}
\label{eq:dm_IP}
\end{subequations}
where we included the equation for $\dot{\tilde\sigma}^*$, because it will
be used lates in the averaging method to second order.
Doing 1st-order averaging, which is equivalent to the RWA, we obtain
\begin{subequations}
\begin{align}
    \dot \Delta &= -\gamma_1(\Delta -\Delta_0)-2\imath\Omega_0 \left[\tilde\sigma e^{-\imath\varphi_0} -\tilde\sigma^*e^{\imath\varphi_0}\right],\\
    \dot{\tilde\sigma} &=-\gamma_2\tilde \sigma-\imath\delta\tilde\sigma
    -\imath\Omega_0\Delta e^{\imath\varphi_0},
\end{align}
\label{eq:dm_rwa}
\end{subequations}
whose fixed-point solutions are given by
\begin{align}
    \tilde\sigma &=-\dfrac{\imath\Omega_0\Delta e^{\imath\varphi_0}}{\gamma_2+\imath\delta}=-\dfrac{\imath\Omega_0\Delta_0(\gamma_2-\imath\delta) e^{\imath\varphi_0}}{\gamma_2^2+\delta^2+4\Omega_0^2\gamma_2/\gamma_1},\\
\Delta &= \frac{\Delta_0}{1+\dfrac{4\Omega_0^2\gamma_2}{\gamma_1(\gamma_2^2+\delta^2)}}.
\end{align}
Next, we proceed to the calculations for the 2nd-order averaging 
corrections in the dynamics of the Bloch equations. 
\subsection{Averaging to 2nd-order}
Again using the notation of Ref.~\cite{bat00} we obtain
\[
\epsilon \tilde f= -\imath\Omega_0\left[
\bea{c}
2\left(\tilde\sigma e^{\imath(2\omega t+\varphi_0)}-\tilde\sigma^* e^{-\imath(2\omega t+\varphi_0)}\right)\\
\Delta e^{-\imath(2\omega t+\varphi_0)}\\
-\Delta e^{\imath(2\omega t+\varphi_0)}\\
\ea
\right]
\]
with the weakly nonlinear transformation $b(t)=y(t)+\epsilon\mathcal{W}(y, t)$,
with $b(t), y(t)\in\mathds{C}^3$,
where  $\mathcal{W}(y, t)$ obeys the differential equation
\beq
\mathcal{W}_t=\tilde f.
\label{eq:Wt2}
\eeq
Hence, upon integration of Eq.~\eqref{eq:Wt2}, we obtain
\beq
\epsilon\mathcal{W}(y, t)=-\frac{\Omega_0}{2\omega}\left[\bea{c}
2\left(\tilde\sigma e^{\imath(2\omega t+\varphi_0)}+\tilde\sigma^* e^{-\imath(2\omega t+\varphi_0)}\right)\\
-\Delta e^{-\imath(2\omega t+\varphi_0)}\\
-\Delta e^{\imath(2\omega t+\varphi_0)}\\
\ea
\right],
\label{eq:W2}
\eeq
in which the constants of integration were set to zero.
The Jacobian matrix of $\tilde f$ is
\[
\epsilon D\tilde f=-\imath\Omega_0
\left[
\bea{ccc}
0 & 2e^{\imath(2\omega t+\varphi_0)} & -2e^{-\imath(2\omega t+\varphi_0)}\\
e^{-\imath(2\omega t+\varphi_0)} & 0 &0\\
-e^{\imath(2\omega t+\varphi_0)} & 0 &0\\
\ea
\right]
\]

\[
\epsilon^2\overline{D\tilde f\mathcal{W}}
=\frac{\imath\Omega_0^2}{\omega}
\left[
\bea{c}
0\\
\tilde\sigma\\
-\tilde\sigma^*
\ea
\right].
\]
Hence, in the second-order averaging we obtain
\begin{subequations}
\begin{align}
    \dot \Delta &= -\gamma_1(\Delta -\Delta_0)-2\imath\Omega_0 \left[\tilde\sigma e^{-\imath\varphi_0} -\tilde\sigma^*e^{\imath\varphi_0}\right],\\
    \dot{\tilde\sigma} &=-\gamma_2\tilde \sigma-\imath\left(\delta-\dfrac{\Omega_0^2}{\omega}\right)\tilde\sigma
    -\imath\Omega_0\Delta e^{\imath\varphi_0}.
\end{align}
\label{eq:dm_2nd_avg}
\end{subequations}
One can readily see that in this approximation the driving phase has no 
effect on the population dynamics, and just adds a constant phase to 
$\tilde\sigma$.

\section{Application to quantum control}
\label{sec:qcontrol}
We now proceed to apply the averaging method to study the action of pulses on
the dynamics of TLS's.
In particular, we are interested in designing $\pi$-pulses to accomplish 
the deepest population inversions.
By examining Eq.~\eqref{eq:dm_2nd_avg}, one sees that the usual area theorem 
\cite{mccall67, mccall67b, allen87} does not work. 
A pulse with central frequency at $\omega=\omega_0$ loses resonance  as the
pulse amplitude increases and, consequently, if the peak amplitude is large
enough, there is substantial detuning and dephasing that population inversion
can no longer be accomplished.
Furthermore, the phase of the pulse may become very relevant in determining 
the effectiveness of the pulse excitation, especially so the shorter the
pulse is, whereas in the RWA or in the averaging approximation to 2nd-order,
it plays no role.
Examining Eqs.~\eqref{eq:dm_2nd_avg} one sees that if the detuning of the
pulse is zero during the driving, then one could achieve population inversion.
Based on this theoretical result, one could devise two strategies: one with chirped
pulses that take into account the amplitude dependent detuning of the
Bloch-Siegert shift, and the other one with pulse shaping instead of chirping.

In the first strategy one could have a chirp on the pulse central frequency
given by $\omega(t)=\omega_0+\Omega_0(t)^2/\omega(t)$, hence the area theorem
could be valid again. 
This should be the case, provided the time variation of $\omega(t)$ is not
too fast.
Next, we will use only the positive root of this quadratic equation,
which is given by
\begin{equation}
\omega(t)=\dfrac{\omega_0+\sqrt{\omega_0^2+4\Omega_0^2(t)}}{2}.
\label{chirp}
\end{equation}
Furthermore, one needs $\Omega_0(t)$ to vary slowly in time compared
with $2\pi/\omega_0$ and $\Omega_0(t)=O(\epsilon)$ and $\dot
\omega(t)=O(\epsilon)$ at most, so that the averaging techniques can be
applied.
One should notice that when TLS's are driven by pulses, Floquet
theory is not applicable, whereas averaging techniques may still be applied
provided the above restrictions are maintained and that the transformation to
a rotating frame is given by 
\[
\sigma (t)=e^{\imath\int_0^t\omega(s)ds}\tilde \sigma(t),
\]
as was discussed in Ref.~\cite{bat04b}.
In order to achieve maximum population transfer, a Gaussian $\pi$-pulse should
be given by
\begin{align}
F(t)&=2\Omega_0 (t)\cos\left(\int_{0}^t\omega(s)ds+\varphi_0\right)\nn\\
&=
2\Omega_0e^{-\frac{(t-t_0)^2}{2{\sigma_0^2}}}\cos(\omega_0 t+\varphi(t)),
\end{align}
where the pulse instantaneous frequency $\omega(s)$ is given by
Eq.~\eqref{chirp} and $\sigma_0=\dfrac{\sqrt{\pi}}{2\sqrt{2}\Omega_0}$ and
$t_0>>\sigma_0$. 

\begin{figure}[h]
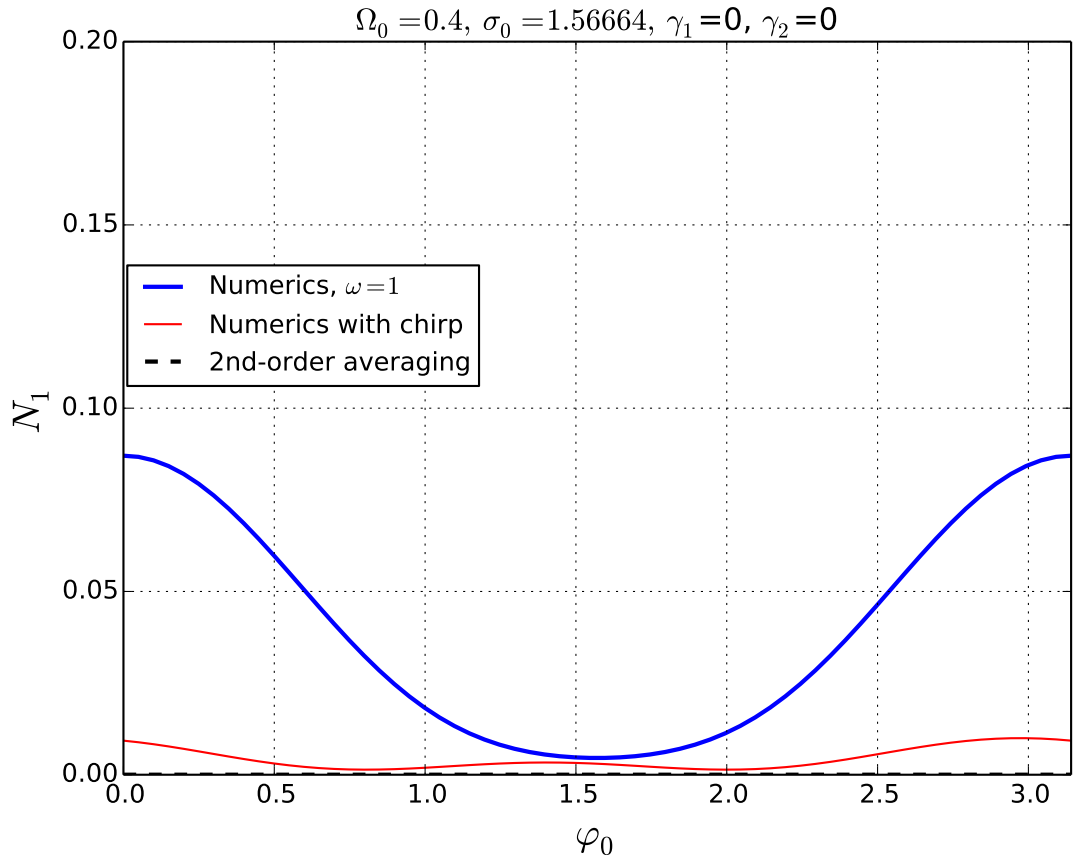

    \centerline{\includegraphics[scale=.8]{{{rabiOscChirpPhaseOmega00.4sigma1.56664omega1}.pdf}}}
        \caption{Population in the ground level after a $\pi$-pulse excitation 
        as a function of pulse carrier phase.
        Before the $\pi$-pulse excitation $N_1=1$.
        The thick blue line depicts the response due to a pulse
        without chirp, with central frequency $\omega=1$, while the thin red 
        line depicts the response to a chirped pulse with the central frequency
        $\omega(t)$ given by Eq.~\eqref{chirp}. 
        The dashed black line is the corresponding 2nd-order averaging result,
        corresponding to an ideal population inversion independent of phase.
        }
\label{fig:chirpedPiPulse}
\end{figure}

\begin{figure}[h]
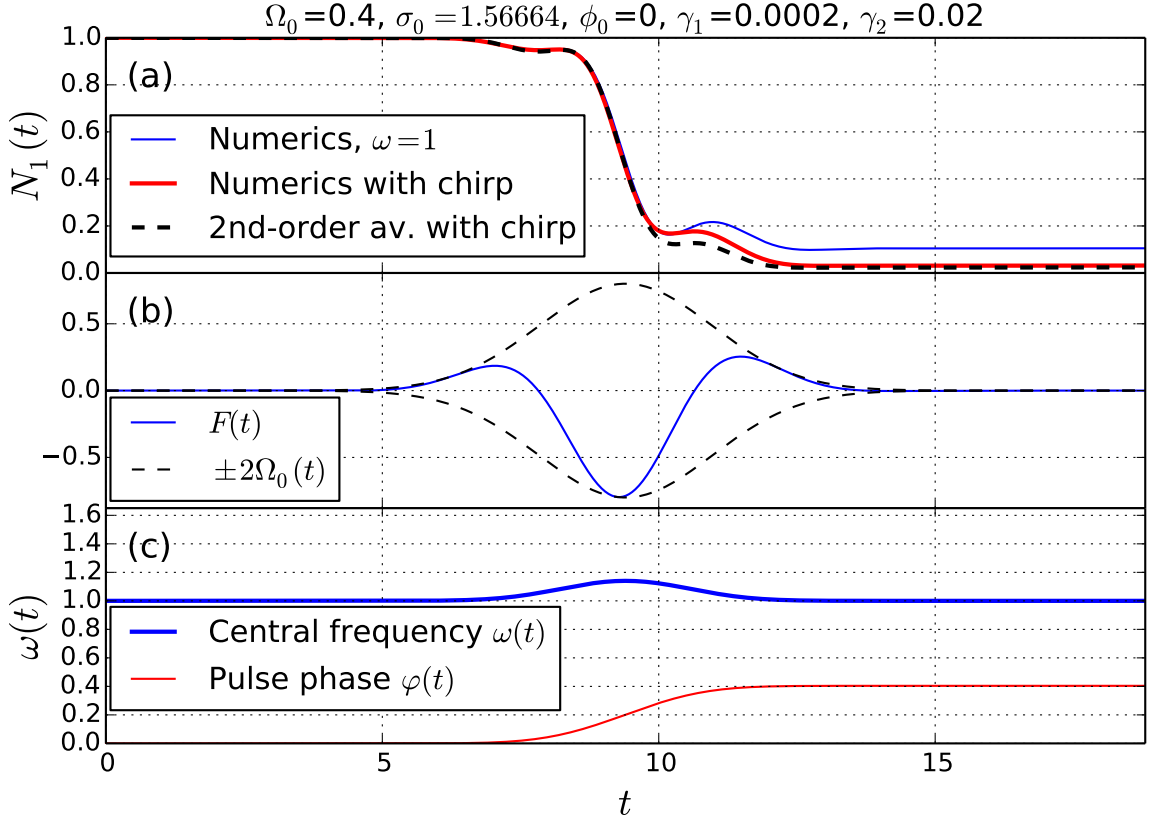

    \centerline{\includegraphics[scale=.8]{{{rabiOscChirpedPulseOmega00.4_1_ga10.0002_ga20.02}.pdf}}}
        \caption{Interlevel transitions driven by $\pi$-pulses with dissipation
        rates in the relaxation time approximation. 
        In frame (a) we show the population of the ground level as a function
        of time. The thin blue line depicts the numerical result of the TLS
        excitation due to a pulse without chirp, with central frequency
        $\omega=1$, the thick red line depicts the numerical results of the
        response to a chirped pulse with the central frequency $\omega(t)$
        given by Eq.~\eqref{chirp}, and the dashed black line represents the
        corresponding averaged equations to second order with the corrections
        of Eq.~\eqref{eq:W} adapted to include chirp. 
        In frame (b) we show the instantaneous chirped pulse along
        with the Gaussian envelope. In frame (c) we plot the central frequency
        and the pulse phase angle as a function of time.}
	\label{fig:pipulseRTA}
\end{figure}

In Fig.~\ref{fig:chirpedPiPulse}, one sees that this strategy yields very
good results, with deep population inversions and little dependance on pulse
carrier phase.
Whereas, $\pi$-pulse excitations with central frequency $\omega=\omega_0$ 
yield results that are very dependent on pulse phase.
This phase dependance becomes a more relevant issue the shorter the pulse is.
The drawback of this method is that it is likely not very easy to design
the chirped pulses experimentally.
In Fig.~\ref{fig:pipulseRTA}, with
the presence of decoherence under the relaxation-time approximation the
proposed strategy of using pulses with chirp yield deeper population invertions
than what would be achieved with a resonant $\pi$-pulse excitation.
One obtains a very good fit for the population inversion with the second-order
averaging result and the numerics with pulse chirping.

The second strategy, though,  involves no chirp, instead it keeps the effective
detuning as small as possible by choosing the pulse maximum amplitude so that
$\Omega_0=\sqrt{\delta\omega}$.
Here we need a near resonance condition of $\delta=\mathcal{O}(\epsilon^2)$,
so that $\Omega_0=\mathcal{O}(\epsilon)$, since $\omega=\mathcal{O}(1)$.
Obviously, for this strategy to be valid, one needs a blue-shifted central
frequency with respect to the transition frequency $\omega_0=1 $ (in our units).
The advantage of this method is that one can have an effective $\pi$-pulse even
with a blue-shifted detuning, what is not possible when the McCall-Hahn's area
theorem is valid, since it is based on the RWA.

\begin{figure}[h]
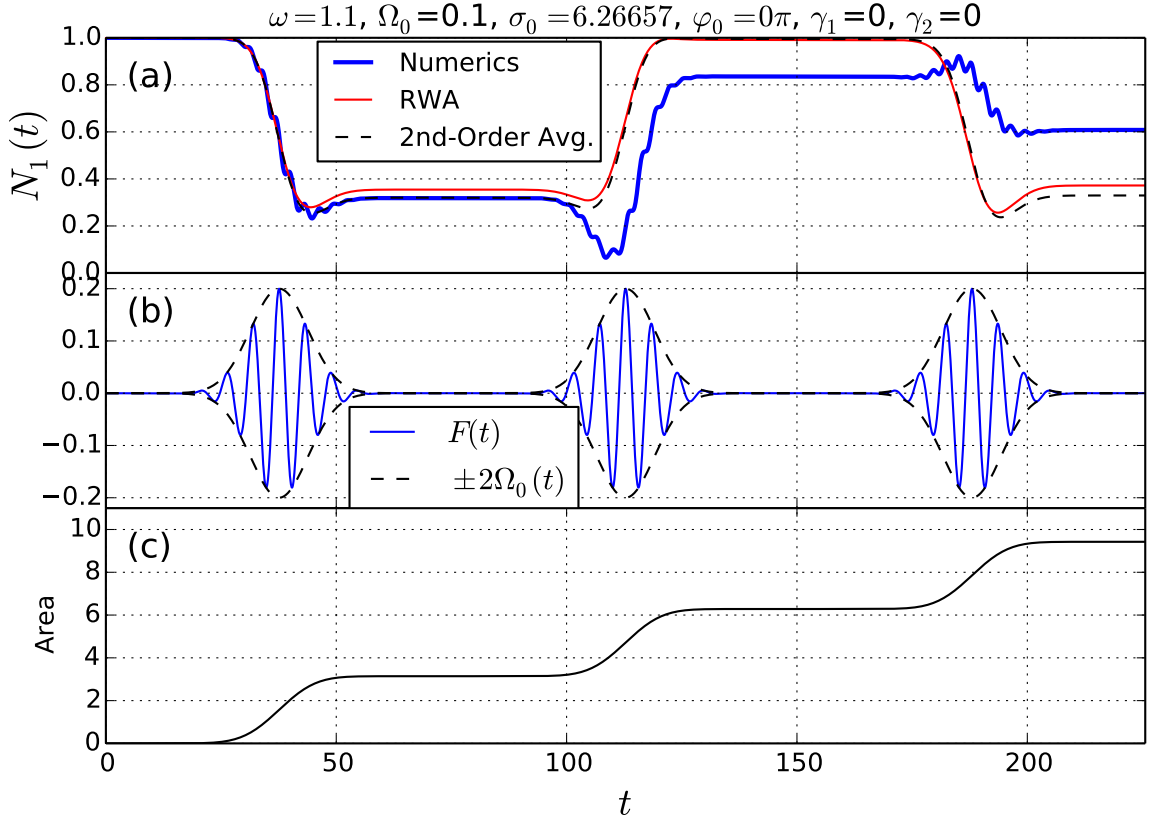

    \centerline{\includegraphics[scale=.8]{{{rabiFlopsOmega00.1om1.1ga10ga20}.pdf}}}
        \caption{Interlevel transitions driven by three $\pi$-pulses. 
        In frame (a) we show the population of the ground level as a function
        of time. 
        The excitations are pulses with carrier frequency at $\omega=1.1$ and
        peak amplitudes with $\Omega_0=0.1$. 
        The width and phase of the pulses are indicated in the figure.
        In frame (b) we show the instantaneous pulse shapes along with the
        $\pi$-pulse Gaussian envelopes. 
        In frame (c) we plot the accumulated area of the pulse envelopes as 
        a function of time, which shows that we actually have $\pi$-pulses.}
\label{fig:pipulses}
\end{figure}

In Fig.~\ref{fig:pipulses}, we show a sequence of three Gaussian $\pi$-pulses
each one with a central frequency of $\omega=1.1$ and  amplitude 
$2\Omega_0=0.2$ acting on the TLS.  
Due to the detuning almost only about 40\% of the ground state population 
is excited after the three $\pi$-pulses.  
Both the RWA and averaging to 2nd-order predict well the behavior of the 
population after the first $\pi$-pulse, but break down afterwards as can be seen in frame (a).  
In frame (b) we show the instantaneous pulse shapes along with their envelopes
and in frame (c) we show the accumulated area of the pulse envelopes as a
function of time to attest that indeed we have $\pi$-pulses.
One sees that if the pulse shape is not well chosen the $\pi$-pulses are not
effective, the RWA and averaging approximation break down, and one loses
quantum control over the qubit.

\begin{figure}[h]
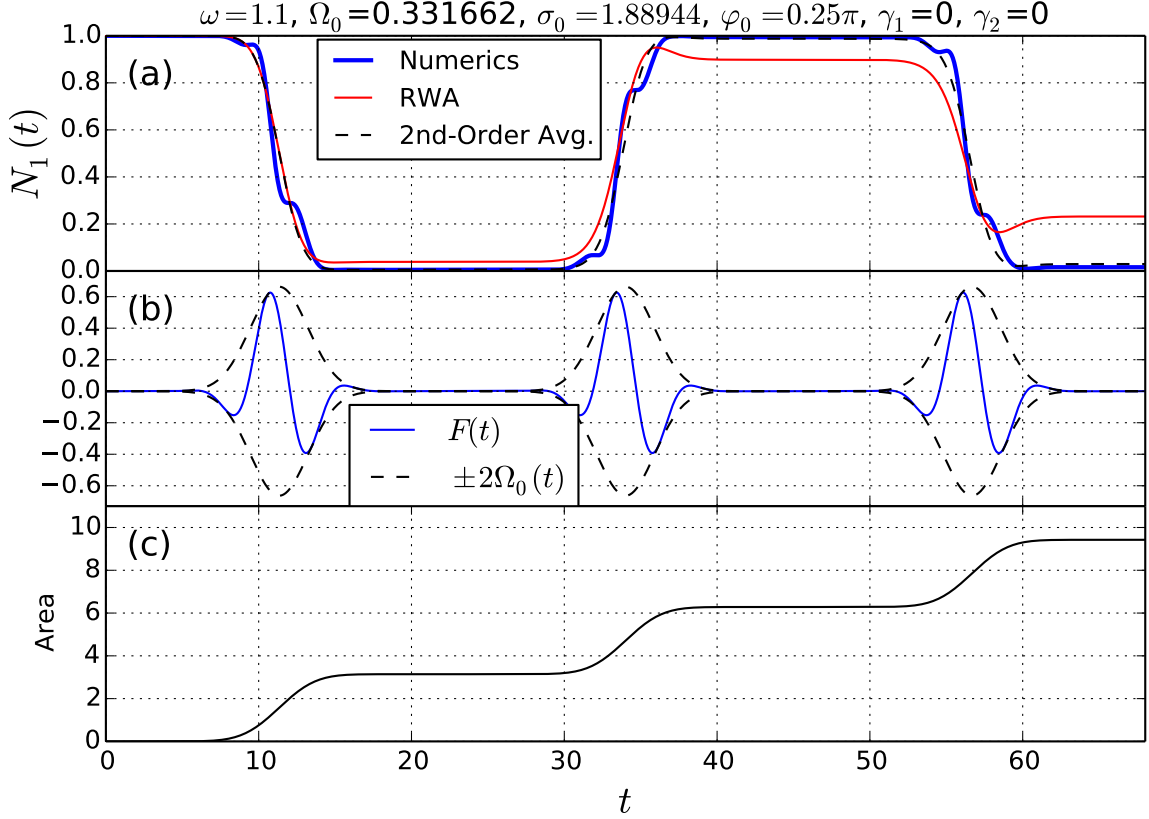

    \centerline{\includegraphics[scale=.8]{{{rabiFlopsOmega00.331662om1.1ga10ga20}.pdf}}}
        \caption{Interlevel transitions driven by three $\pi$-pulses. 
 In frame (a) we show the population of the ground level as a function
        of time. The thick blue line depicts the numerical integration result. 
        One sees that complete population inversions are achieved, despite
        the fast dynamics, in which the RWA result (red line) is no longer
        valid. On the other hand, the results of the averaging method to
        second-order (dashed line) are still very useful in designing effective  $\pi$-pulses.
        In frame (b) we show the instantaneous pulse shapes with central
        frequency at $\omega=1.1$ along  with the $\pi$-pulse Gaussian
        envelopes with peak amplitude with $\Omega_0=\sqrt{\delta\omega}$. 
        One sees that one gets about only one and a half cycles per pulse. 
        In frame (c) one sees the accumulated area of the pulse envelopes as 
        a function of time, at each pulse the area increases by $\pi$.
}

\label{fig:pipulsesb}
\end{figure}

In Fig.~\ref{fig:pipulsesb}, we again show a sequence of three Gaussian
$\pi$-pulses, each one with a central frequency of $\omega=1.1$ and envelope
amplitude with $\Omega_0=\sqrt{\delta\omega}=0.331662$ acting on the TLS.
One sees that due to the detuning, after three $\pi$-pulses the RWA (red line)
breaks down yielding less than 80\% of population inversion.
Whereas the averaging to 2nd-order still got approximately 98\% of population 
inversion.
With this value of pulse envelope amplitude, one achieves nearly 99\% of
population inversion by numerically integrating the density matrix equations of
motion in Eqs.~\eqref{eq:dm}.
This indicates that the averaging method to second-order is a good predictor
of the behavior of the population transfer between the ground and
excited levels when our pulse shaping strategy is implemented.
In frame (b) we show the instantaneous pulse shapes along with their envelopes
and in frame (c) we show the accumulated area of the pulse envelopes as a
function of time to attest that indeed we have $\pi$-pulses.

\begin{figure}[h]
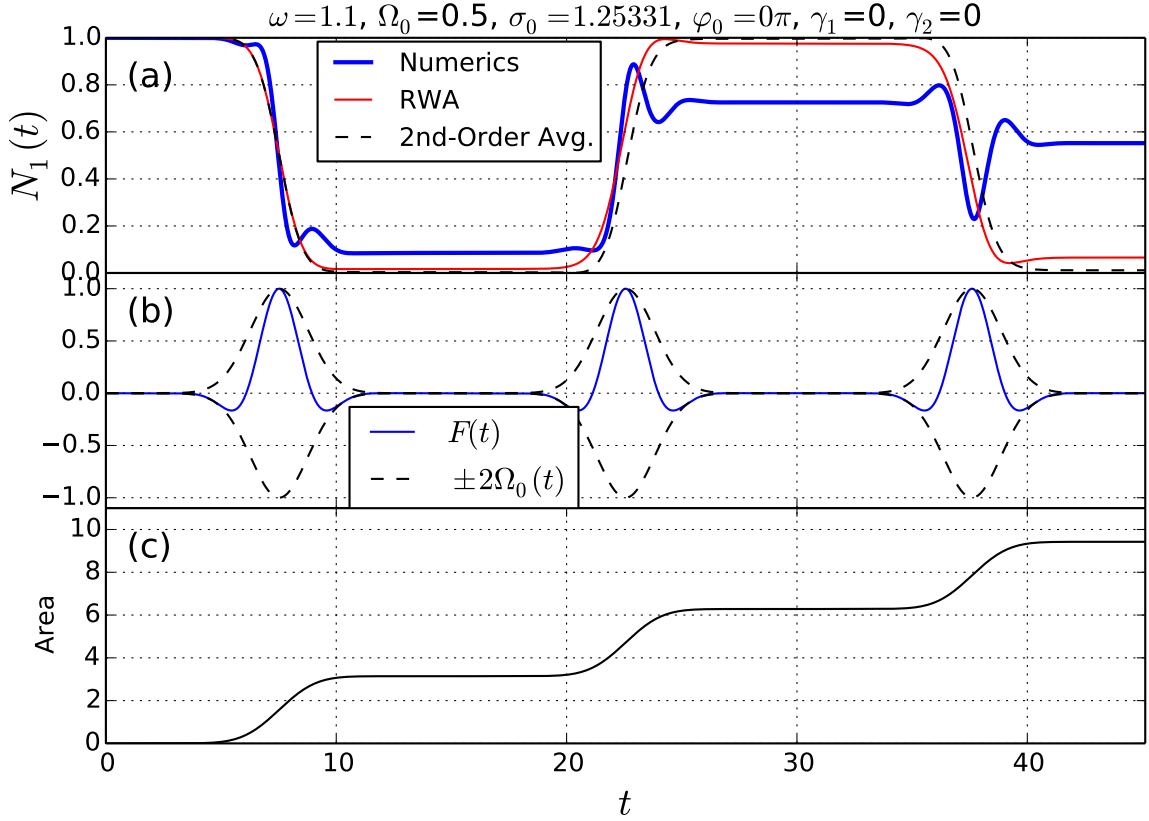

    \centerline{\includegraphics[scale=.8]{{{rabiFlopsOmega00.5om1.1ga10ga20}.pdf}}}
        \caption{Interlevel transitions driven by $\pi$-pulses. 
        In frame (a) we show the population of the ground level as a function
        of time. 
        The excitations are pulses with carrier frequency at $\omega=1.1$ and
        peak amplitudes with $\Omega_0=0.5$. 
        The width and phase of the pulses are indicated in the figure.
        The RWA and the averaging results both break down here. 
        In frame (b) we show the instantaneous pulse shapes along with the
        $\pi$-pulse Gaussian envelopes. 
        In frame (c) we plot the accumulated area of the pulse envelopes as 
        a function of time, which verifies that we actually have $\pi$-pulses.}
\label{fig:pipulsesc}
\end{figure}

In Fig.~\ref{fig:pipulsesc}, we show a sequence of three Gaussian $\pi$-pulses
each one with a central frequency of $\omega=1.1$ and  amplitude 
of twice $\Omega_0=0.5$ acting on the TLS.  
Due to the detuning almost only about 44\% of the ground state population 
is excited after the three $\pi$-pulses.  
Both the RWA and averaging to 2nd-order break down as can be seen in frame (a).
In frame (b) we show the instantaneous pulse shapes (with zero phase) along
with their envelopes
and in frame (c) we show the accumulated area of the pulse envelopes as a
function of time to attest that indeed we have $\pi$-pulses.
Again, one sees that if the pulse shape is not well chosen the $\pi$-pulses are
not effective, the RWA and averaging approximation break down, and one loses
quantum control over the qubit.

\begin{figure}[h]
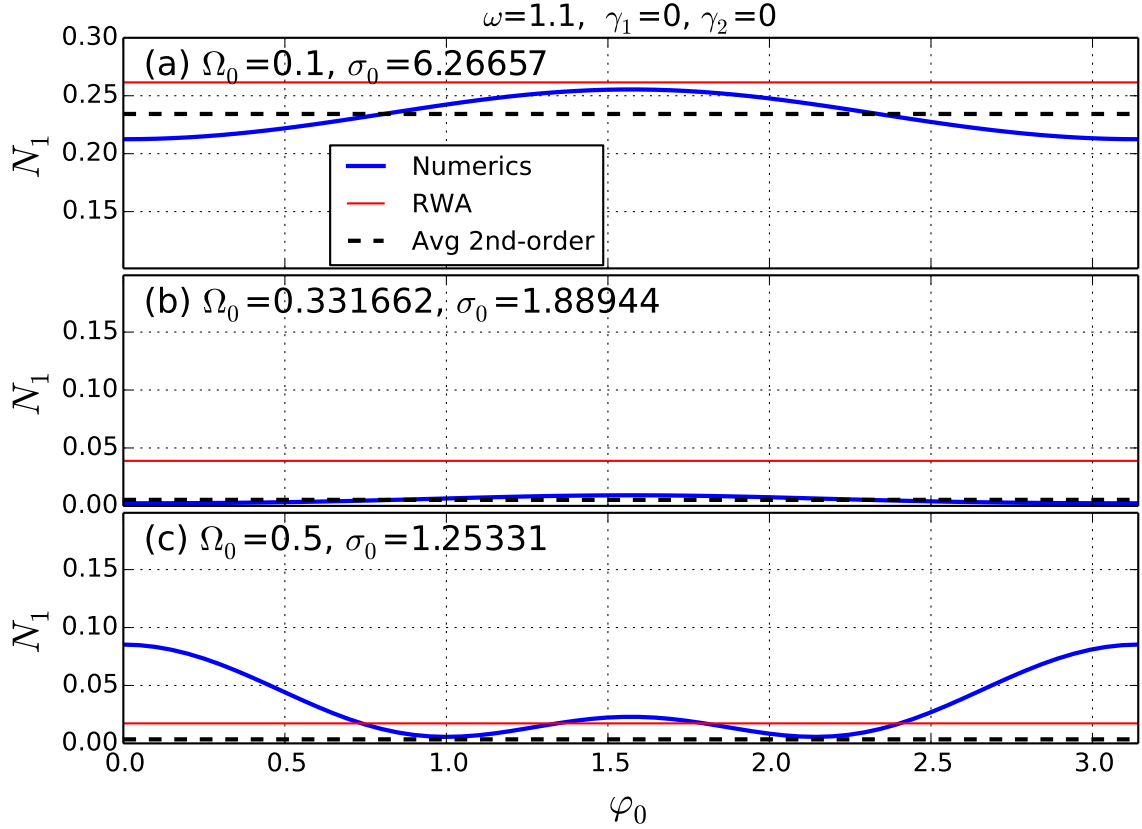

    \centerline{\includegraphics[scale=.8]{{{rabiOscPhasesOmega00.5sigma_01.25331omega1.1}.pdf}}}
        \caption{Ground level population as a function of pulse phase after 
        various $\pi$-pulse excitations with carrier frequency at $\omega=1.1$. 
        Note that before the excitation $N_1=1$ and that all pulse
        envelopes are Gaussian with amplitudes and widths indicated in each
        frame.
        In frame (a) we have dephasing due to the Bloch-Siegert shift and
        thus the outcome of the $\pi$-pulse is not a complete inversion and
        is very phase dependant.
        In frame (b) when the pulse amplitude is 2$\Omega_0=2\sqrt{\delta\omega}
        \approx0.6632$ the pulse is effectively at resonance, with nearly no
        dephasing due to the Bloch-Siegert shift.
        One sees that for this value of amplitude, one gets nearly complete
        inversion and very little phase dependance. This indicates that the 
        pulse  shaping procedure based on  Eqs.~\eqref{eq:dm_2nd_avg}, i.e. by
        keeping the detuning as low as possible, is very effective.
        In frame (c) we again have dephasing due to the Bloch-Siegert shift and
        thus the outcome of the $\pi$-pulse is very phase dependant.
}
\label{fig:pulsePhases}
\end{figure}

The results presented in Fig.~\ref{fig:pulsePhases} show the amount of
population remaining in the ground state after a Gaussian $\pi$-pulse
excitation is applied to the TLS.
The shapes of the pulses in frames (a)-(c) are equal to the shapes of the
the pulses in Figs. \ref{fig:pipulses}-\ref{fig:pipulsesc}, respectively.
In frame (a) there is no complete population inversion (under 80\%) and
the outcome of the pulse excitation depends considerably on the phase of pulse.
Whereas, in frame (b), we notice that our pulse shaping strategy to generate 
the deepest inverstions works very well. 
Here, due to the Bloch-Siegert shift, the pulse is effectively in resonance,
hence it is a $\pi$-pulse.
We obtain very good fit with the averaging and the numerical results with
nearly complete inversion independent of pulse phase, what make performing 
quantum control of qubits is easier. 
In frame (c), since one gets again out of resonance due to the Bloch-Siegert
shift, not only there is no complete population inversion, but there is also
strong influence of the phase of the pulse on the outcome of the $\pi$-pulse
excitation. 

\section{Conclusion}
\label{sec:conclusion}
Here we showed that perturbation techniques based on the averaging method
can efficiently describe the unitary (or near unitary) time evolution
of TLS's under coherent excitation.
With the averaging method to second-order approximation, we went beyond the RWA
in analyzing the dynamics of such a fundamental process in quantum mechanics.
The numerical time-evolution of the ground state population of the TLS under cw driving had a far better fit with the averaging method to 
2nd-order approximation than with the RWA time evolution.
Basically, this better approximation was achieved by eliminating the dephasing
that occurs in  the RWA.
By going to second-order in the averaging method, we obtained  the
Bloch-Siegert shift of the TLS resonant frequency, in which the
resonant frequency increases quadratically with the driving field amplitude.
This frequency shift is the main cause of the dephasing of the ground state
population time evolution observed in the RWA.
 
Although, the proposed method is not exact as Floquet theory is in the
case of cw driving of TLS's, it is far easier to implement than
the latter theory.
Furthermore, it can still be applied even when Floquet theory can no longer be
applied, such as when the TLS is driven by short pulses.

We verified that the Bloch-Siegert frequency shift that occurs in cw driving,
also has important influence on pulse-driven TLS's.
This shift basically implies that the well-known area theorem of McCall and
Hahn breaks down when the Rabi frequency of pulse-driven TLS's
becomes approximately of the order ($>40$\%) of the transition frequency.
By numerically integrating the optical-Bloch equations we note that, for 
intense and short $\pi$-pulse excitations,
consistently more population inversion is achieved and with less phase
dependance if the pulse central frequency has a chirp or
the pulse amplitude is chosen so as to minimize the detuning effect of the 
Bloch-Siegert shift.
 
The present approaches of designing $\pi$-pulse shapes guided by the averaging
method could be used in experimentally generating shorter $\pi$-pulses that
achieve considerably more population inversion than with $\pi$-pulses without
chirp or with the wrong shape.
Our results may prove to be very helpful in speeding up control schemes of
quantum qubits in quantum computation, since it minimizes the 
effects of pulse phase and decoherence on the effectiveness of the $\pi$-pulses.
If one only achieves an imperfect population inversion, such as around 95\% of
inversion, after a few $\pi$-pulses one could lose the control over the qubits.
The control of qubits for the realization of quantum gates is of fundamental
importance in quantum computation \cite{nielsen2010quantum},
as they are usually realized by sequences of laser pulses in physical systems,
such as, for instance, ion traps \cite{monz2009realization, schmidt2003realize}.
It is important to mention that our methods are not restricted to the design of
just the Gaussian-shaped $\pi$-pulses shown here, but they may be realized with
other types of pulses used in the implementation of quantum gates.
Furthermore, our results may prove to be useful for increasing the photocurrent
in quantum dot (QD) devices \cite{zrenner02} even when the pulse carrier
frequency is not in resonance with the interlevel transition frequency of the
QD.

Finally, we would like to point out that the fairly straight forward and
powerful capabilities of the averaging method, such as those used here, should
allow it to have a place among the usual time-dependent perturbative methods
taught in standard text books of quantum mechanics.
%\bibliography{/home/adriano/research/artigos/gen2}%>> bibliography data
%\bibliographystyle{apsrev4-1}   %>>>> makes bibtex use prbbib.bst

%figures
\newpage
\FloatBarrier

\begin{figure}[h]
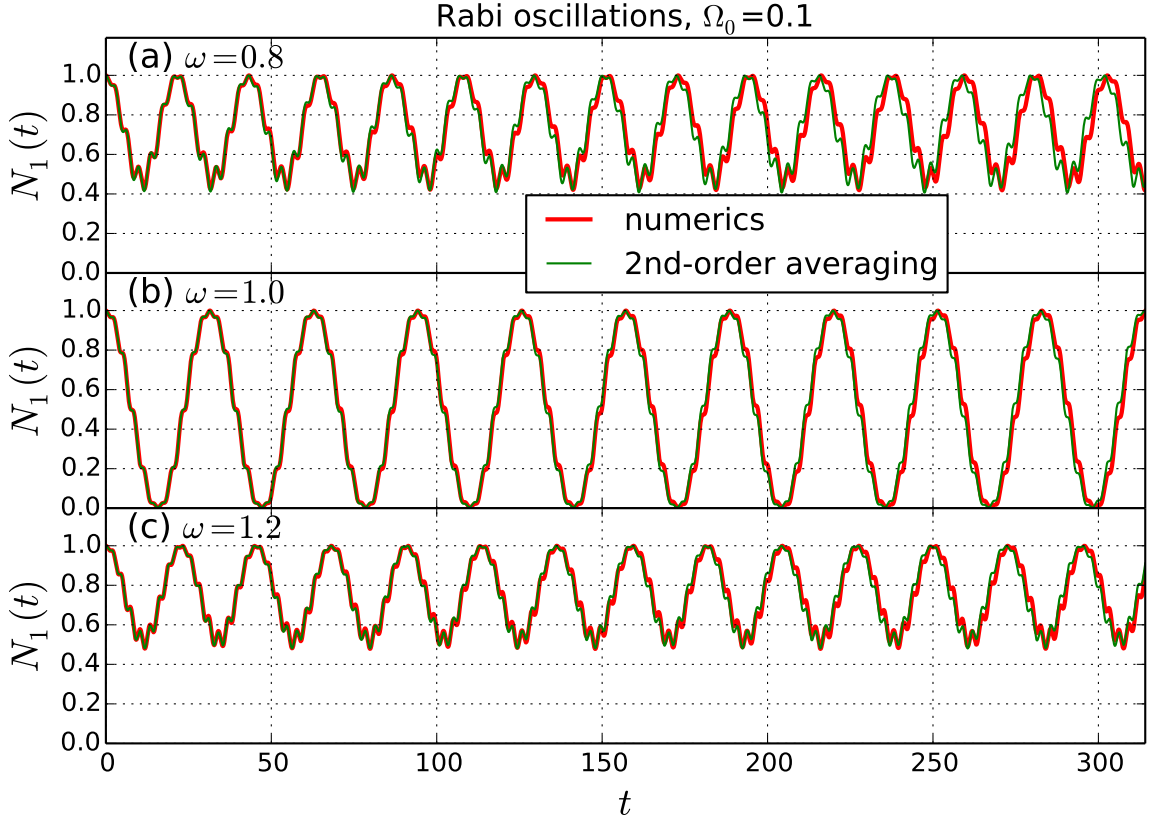

    \centerline{\includegraphics[scale=0.8]{{{rabiOsc2nd_avg_Omega0.1}.pdf}}}
        \caption{Comparison of the ground state population time evolution as
        predicted by the 2nd-order averaging approximation of Eqs.~\eqref{eq:rabi_avg2} with the corrections of Eq.~\eqref{eq:W} and the result given
        by numerical integration of Eqs.~\eqref{eq:2level}.
        One can still see some dephasing, but it is far less pronounced than in
        the RWA results of Fig.~\ref{fig:Rabi_flopRWAa}.}
        \label{fig:Rabi_flop_avg2}
\end{figure}
\FloatBarrier
\FloatBarrier
\FloatBarrier

\FloatBarrier
\FloatBarrier

\end{document}